# Skyrmion-skyrmion and skyrmion-edge repulsions in skyrmion-based racetrack memory

Xichao Zhang[1], G. P. Zhao[1, 2, *], Hans Fangohr[3], J. Ping Liu[2, 4], W. X. Xia[2], J. Xia[1], F. J. Morvan[1]

[1]College of Physics and Electronic Engineering, Sichuan Normal University, Chengdu 610068, China, [2]Key Laboratory of Magnetic Materials and Devices, Ningbo Institute of Material Technology & Engineering, Chinese Academy of Sciences, Ningbo 315201, China, [3]Engineering and the Environment, University of Southampton, Southampton SO17 1BJ, United Kingdom, [4]Department of Physics, University of Texas at Arlington, Arlington, Texas 76019, USA

Magnetic skyrmions are promising for building next-generation magnetic memories and spintronic devices due to their stability, small size and the extremely low currents needed to move them. In particular, skyrmion-based racetrack memory is attractive for information technology, where skyrmions are used to store information as data bits instead of traditional domain walls. Here we numerically demonstrate the impacts of skyrmion-skyrmion and skyrmion-edge repulsions on the feasibility of skyrmion-based racetrack memory. The reliable and practicable spacing between consecutive skyrmionic bits on the racetrack as well as the ability to adjust it are investigated. Clogging of skyrmionic bits is found at the end of the racetrack, leading to the reduction of skyrmion size. Further, we demonstrate an effective and simple method to avoid the clogging of skyrmionic bits, which ensures the elimination of skyrmionic bits beyond the reading element. Our results give guidance for the design and development of future skyrmion-based racetrack memory.

Magnetic skyrmions are topologically stable nanoscale magnetization configurations, which have been discovered in certain magnetic bulks, films and nanowires [1-13]. Individual skyrmions have already been created using a spin-polarized current experimentally [11]. As for skyrmion lattices, they have been observed in MnSi, FeCoSi and other B20 transition metal compounds [1-10] as well as in helimagnetic MnSi nanowires [12, 13]. Two-dimensional skyrmion crystals in nanopatterned Co/CoPt bilayers [14] and spontaneous skyrmion ground states in Co/Ru/Co nanodisks [15] have also been predicted numerically. Further, numerical studies have demonstrated the current-driven motion of skyrmions in constricted geometries, along with the creation of skyrmions in nanodisks and nanowires with spin-polarized currents [16-20]. An effective particle model has been derived analytically, including repulsive skyrmion-skyrmion interactions, interaction with defects, and the role of the Magnus force [21].

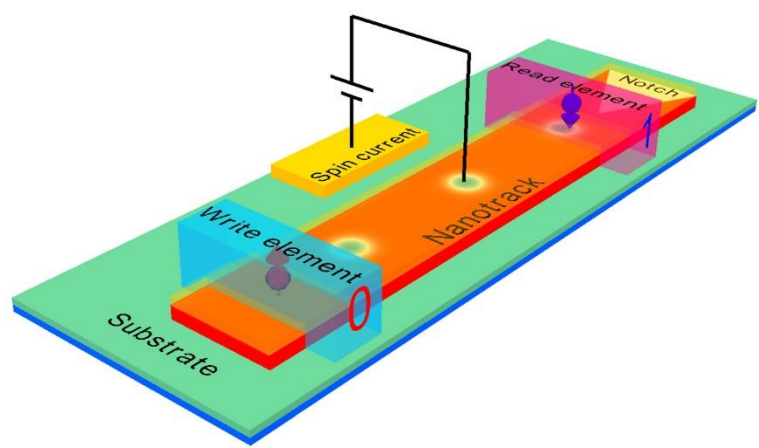

**Figure 1. Schematic of typical vertical-current-driven horizontal skyrmion-based racetrack memory.** Skyrmion-based racetrack memory storage device moves the skyrmions along the racetrack in one direction only. The reading element can be positioned at one end of the racetrack. The skyrmions are annihilated upon moving them across the reading element but their corresponding information is read into one or more memory devices (e.g., built-in CMOS circuits).

Magnetic skyrmions are promising candidates as bit carriers for future data storage and logic devices using extremely low energy, with their motion driven by small electrical currents [10, 22-26]. They could be used in skyrmion-based racetrack memory (RM), as shown in Fig. 1, which stores information in the form of skyrmions in magnetic nanowires or nanostrips (*i.e.* "racetracks"). In contrast to domain wall-based RM [27-29], where data are encoded as a pattern of magnetic domains along a portion of the racetrack, skyrmion-based RM can work with much lower current density (~

$10^6$ A/m$^2$) than that needed for domain wall-based RM ($\sim 10^{12}$ A/m$^2$) [10, 18, 20, 24, 25, 29-31]. Skyrmion-based RM combines non-volatile and energy-efficient qualities along with ultra-high density and negligible ohmic heating [12, 18, 20, 29, 32], attracting widespread attention.

However, significant challenges need to be overcome such as driving and annihilation processes of skyrmionic bits on the racetrack [32]. In this article, a micromagnetic numerical approach has been used to study the impacts of the skyrmion-skyrmion repulsion [21] and the skyrmion-edge repulsion [19-21, 33] for skyrmions in a magnetic racetrack geometry. We study the reliable and practicable spacing between consecutive skyrmionic bits on the racetrack that are driven by vertically injected spin-polarized currents, and demonstrate the ability to adjust that spacing. The behaviors of the skyrmionic bit chain are discussed at the end of the racetrack, where skyrmionic bits are found to squeeze together, compressing the skyrmions and reducing their size. Finally, we propose and test a technological method to avoid the clogging of moving skyrmionic bits at the end of the racetrack by nanolithography, which will eliminate skyrmionic bits efficiently from the racetrack at low current densities. Our goal is to provide guidance for the design and development of future skyrmion-based RM.

## Results

For the purpose of studying the reliable and optimal spacing between consecutive skyrmionic bits, we first discuss the repulsion between two skyrmions on the racetrack without the driving spin current. We initially create two 5-nm-diameter skyrmions with spacings of 30 nm, 52 nm, 57 nm, 62 nm respectively, on the center of a 400-nm-long by 40-nm-wide racetrack by means of an injected spin-polarized current pulse perpendicular to the racetrack plane, which can be done by a movable metal-coated tip or a write element over the racetrack. The details of creating skyrmion via the injection of spin-polarized current perpendicular to the plane are shown in Methods and Ref. 20. When the pulse stops, two skyrmions have been created, one at each tip. Due to their repulsion, the spacing between the skyrmions as well as their sizes change with time. We define the spacing between two consecutive skyrmions $d$ as the distance between their centers (see inset in Fig. 2) and the skyrmion size $r_s$ is defined as the diameter of the circle where m$_z = 0$ (see inset in Fig. 3).

In Ref. 21, the repulsive force $F_{ss}$ between two skyrmions is described by $K_1\left(\frac{d\sqrt{HA}}{D}\right) \times \left(\frac{A}{t}\right)$, where $K_1$ is the modified Bessel function and $H$ is the perpendicular applied field. Hence, we can estimate the skyrmion repulsion force $F_{ss}$ by calculating $K_1\left(\frac{d\sqrt{H_k A}}{D}\right) \times \left(\frac{A}{t}\right)$, where $H_k$ is the perpendicular anisotropy field instead of $H$.

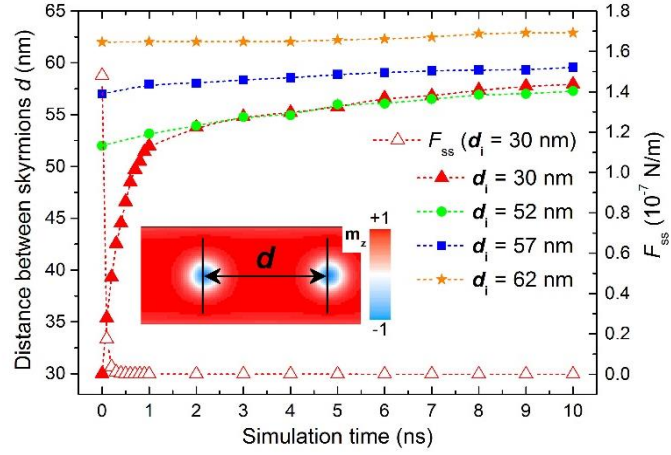

**Figure 2. Distance *d* and repulsive force *F*ₛₛ between two skyrmions on the racetrack as functions of the simulation time.** The initial spacing between the two skyrmions is defined by $d_i$, while $F_{ss}$ stands for the repulsive force between them. The inset shows how we define the distance *d* between two skyrmions where the out-of-plane component of magnetization is presented by the color scale. Symbols are numerical results and dashed lines are used to guide the eyes. The color scale has been used throughout this paper.

Figure 2 shows the skyrmion separation distance *d* as well as the skyrmion repulsion force $F_{ss}$ for different initial spacing $d_i$ *versus* the simulation time. It can be seen that, for the case of initial skyrmion spacing $d_i = 30$ nm, the distance *d* increases rapidly within the first 1 ns from 30 nm to 52 nm, then slowly increases to 58 nm at t = 10 ns. At the same time, $r_s$ increases from 5 nm to ~ 11 nm within the first 1 ns. The corresponding $F_{ss}$ dramatically decreases in first 0.5 ns from ~ $10^{-7}$ N/m to ~ $10^{-10}$ N/m, and then slowly reduces to ~ $10^{-12}$ N/m at 10 ns. From the case of $d_i = 52$ nm ($F_{ss} \approx 10^{-11}$ N/m), we can see that the skyrmion-skyrmion repulsion is weak but still can slowly push the skyrmions away from each other from 52 nm to 57 nm in 10 ns with a rate (0.5 m/s) lower than that of $d_i = 30$ nm from 30 nm to 58 nm (2.8 m/s). When $d_i = 57$ nm ($F_{ss} \approx 10^{-12}$ N/m), the rate of increase of *d* is much lower (0.3 m/s), while when $d_i = 62$ nm ($F_{ss} \approx 10^{-13}$ N/m) the rate of increase of *d* is only 0.1 m/s, which results in motion that is not detectable for the timescales of the simulations carried out here, and we choose this rate as one where the distance between skyrmions stays practically constant. This is fulfilled for $d_i > 62$ nm ($F_{ss} < 10^{-13}$ N/m).

Material imperfections – both geometry and impurities[19] – can suppress the slow relative motion between consecutive skyrmions, as can skyrmions further to the left and right in the chain which are constraint in their movements. Therefore, under the assumption of the typical material parameters used here and in Ref. 20, $d_i \sim > 62$ nm would be an ideal initial spacing for writing consecutive skyrmionic bits as well as the identification spacing for reading consecutive skyrmionic bits (*i.e.*, bit length). And $d_i \sim > 57$ nm can also be regard as reliable and practicable initial spacing for consecutive skyrmionic bits, which has been verified in Ref. 20, where $d_i = 60$ nm is employed. We can relate these distances to the material's Dzyaloshinskii-Moriya interaction (DMI) helix length $L_D$ [17] ($L_D = 4\pi A/|D|$, which equals 62.8 nm in our case) to find the ideal and practical spacings are $\sim L_D$ and $\sim 0.9\ L_D$, respectively.

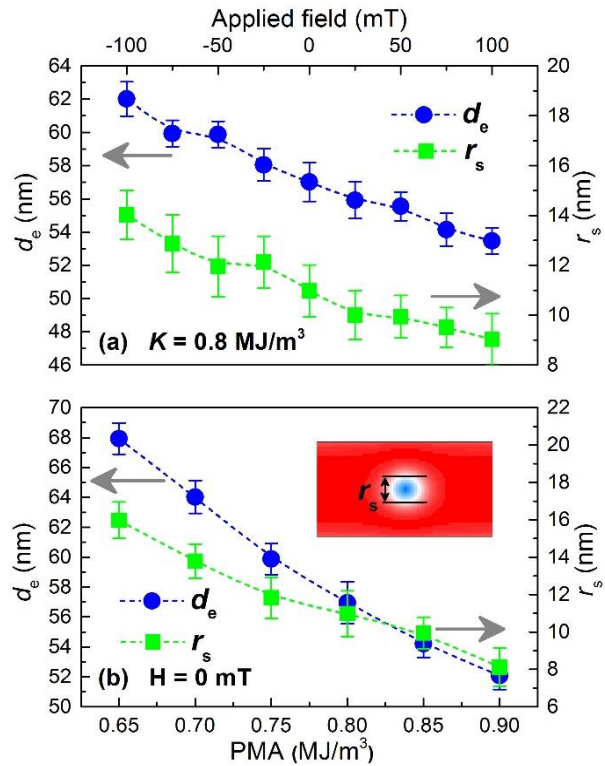

**Figure 3. Effect of different (a) applied fields and (b) perpendicular magnetic anisotropies (PMA) on the equilibrium spacing $d_e$ between consecutive skyrmions and equilibrium skyrmion size $r_s$ on 40-nm-wide racetrack.** The inset shows the definition of the size of a skyrmion as the diameter of a circle at which $m_z = 0$. Symbols are numerical results and dashed lines are used to guide the eyes.

We next investigate the effects of applied field, anisotropy and racetrack geometry on the equilibrium distance of two consecutive skyrmionic bits as well as their size. We initially create two 5-nm-diameter skyrmions with spacing of 30 nm in the center of a 400-nm-long by 40-nm-wide racetrack. At the same time, we apply an external magnetic field perpendicular to the racetrack plane or tune the PMA to adjust the equilibrium distance $d_e$. We measure the relaxed $r_s$ in this experiment after 1 ns and $d$ at 10 ns where the repulsion is virtually zero. The latter can be regarded as the equilibrium distance $d_e$, *i.e.*, the reliable initial spacing for consecutive skyrmionic bits.

Figure 3 shows the equilibrium distance $d_e$ of the two consecutive skyrmionic bits on the 40-nm-wide racetrack under different applied fields or anisotropies. It can be seen from Fig. 3a that the applied field perpendicular to the plane and opposite to the direction of the magnetization in the core of skyrmions (*i.e.* positive values) can marginally reduce the relaxed skyrmion size $r_s$ as well as the equilibrium skyrmion distance $d_e$. It is in good agreement with the findings on single skyrmions in nanodisks under a magnetic field, as shown in Refs. 20 and 17. Under an applied field of +100 mT, $d_e$ decreases from 57 nm to 53 nm. Figure 3b shows $d_e$ and relaxed $r_s$ *versus* the PMA. We can see that both $d_e$ and relaxed $r_s$ decrease approximately linearly with increasing PMA. When the PMA increases from 0.7 MJ/m$^3$ to 0.9 MJ/m$^3$, $d_e$ decreases from 64 nm to 52 nm, and relaxed $r_s$ decreases from 14 nm to 8 nm. Note that if the PMA or applied field is larger than a critical value, the skyrmions will be unstable and cannot exist in the confined geometry, as shown in Refs. 20 and 17.

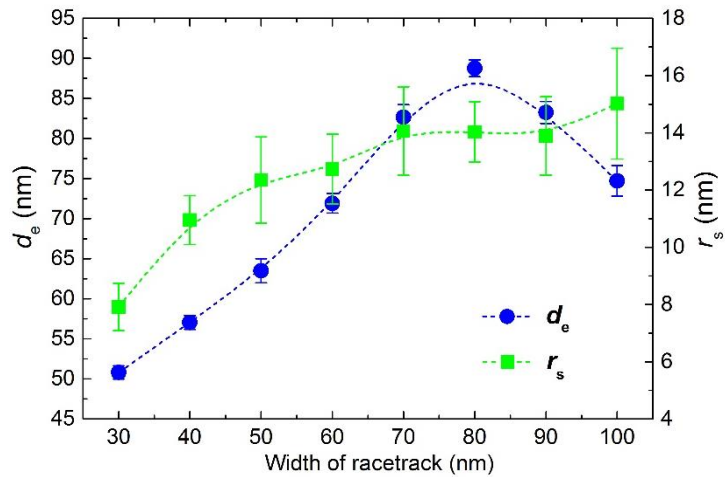

**Figure 4. Skyrmion size $r_s$ and equilibrium spacing between consecutive bits $d_e$ as a function of racetrack width.** Symbols are numerical results and dashed lines are used to guide the eyes.

Figure 4 shows the equilibrium skyrmion distance $d_e$ and the relaxed skyrmion size $r_s$ as functions of the racetrack width. It can be seen that $d_e$ rapidly goes up from ~ 50 nm to ~ 90 nm with the width increasing from 30 nm to 80 nm, then decreases to 75 nm when the width continually increases to 100 nm. In the meantime, $r_s$ increases from 8 nm to 12 nm swiftly when the width increases from 30 nm to 50 nm, and then slowly increases to around 14 nm as the width rises to 100 nm. We can note that when the width of the racetrack is smaller than a certain threshold (~ 20 nm in our simulation), the skyrmionic bits are not stable on the racetrack, which may quickly evolve to domain walls. In Ref. 17, it has also been numerically demonstrated that stable skyrmions only exist in square elements with edge length larger than a given threshold.

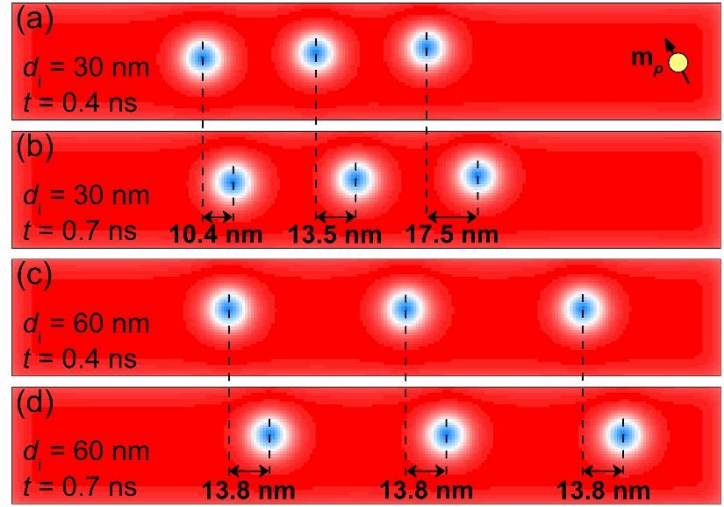

**Figure 5. Vertical-current-driven motion of a skyrmionic bit chain with initial spacing of 30 nm on the 40-nm-wide racetrack at (a) $t$ = 0.4 ns, (b) $t$ = 0.7 ns, and motion of a chain with initial spacing of 60 nm at (c) $t$ = 0.4 ns, (d) $t$ = 0.7 ns.** The current density $j = 5 \times 10^{10}$ A/m², polarization $P = 0.4$ and DMI constant $D = 3 \times 10^{-3}$ J/m². The direction of the electron polarization $\mathbf{m}_p$ is shown in the upper right corner, which is in the film plane and makes a 30-degree angle to the positive $y$-direction.

To further investigate the reliable spacing between consecutive skyrmionic bits in racetrack memory, we performed simulations of vertical-current-driven motion of skyrmionic bit chains with different initial spacing. Figure 5 shows the skyrmionic bit chains with different $d_i$ driven by vertically injected spin-polarized currents on 40-nm-wide by 240-nm-long racetracks. In simulations of vertical-current-driven skyrmionic bits, the currents are perpendicularly injected into the plane of the racetrack with a fixed density of $j = 5 \times 10^{10}$ A/m². The direction of the electron polarization ($P$ = 0.4) is in-plane and makes a 60 degrees angle with the length-axis of the racetrack as shown in Fig. 5, which ensures that skyrmions are driven by the current moving along the racetrack towards right but without pressing strongly on the upper track border [20, 34, 35]. We have found in our simulations that using 90 degrees, results in skyrmions being destroyed at the upper track border (see Supplementary Movies 1 and 2).

In the case of current-driven skyrmionic bit chain with $d_i = 30$ nm, the skyrmion-skyrmion repulsion leads to different velocities until the spacing increases to $d_e$. From t = 0.4 ns to t = 0.7 ns, the right skyrmionic bit moves 17.5 nm ($v = 58$ m/s) while the left skyrmionic bit only moves 10.4 nm ($v = 35$ m/s). But for the case of current-driven skyrmionic bits chain with initial skyrmion distance $d_i = 60$ nm, as shown in Fig. 5c-d, within the same time frame, both skyrmionic bits move 13.8 nm with a steady velocity of 46 m/s.

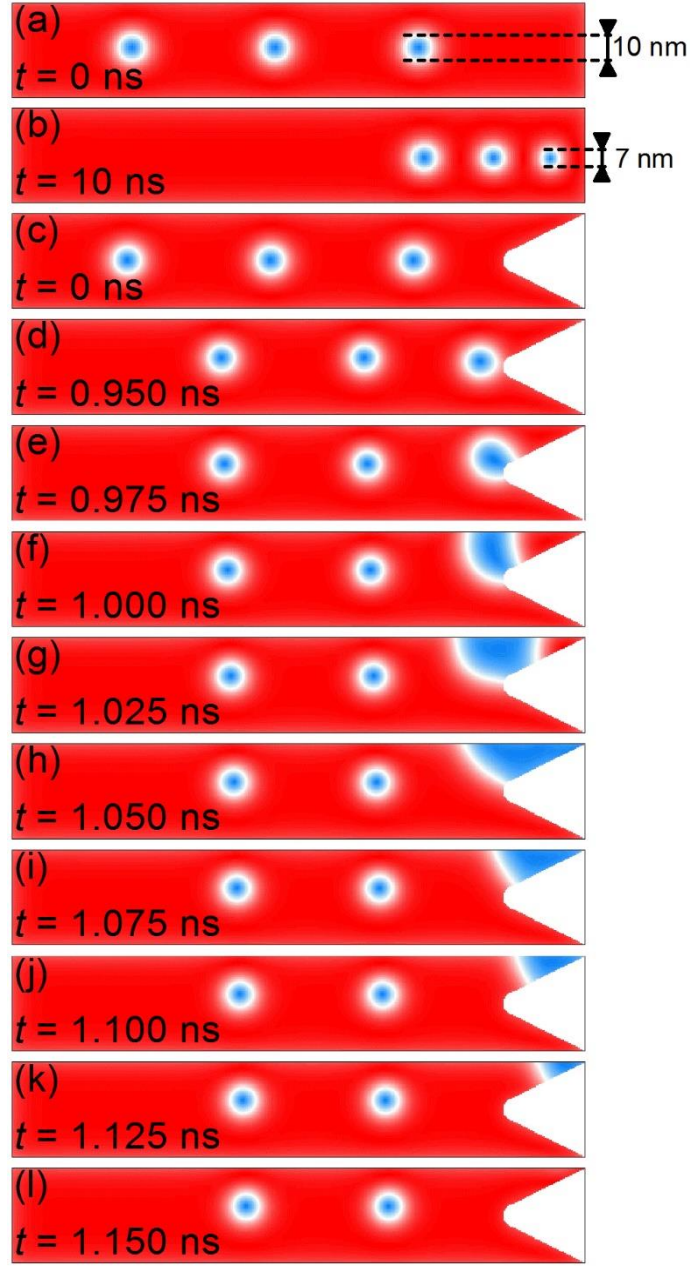

**Figure 6. Vertical-current-driven motion of a skyrmionic bit chain at the end of the 40-nm-wide racetrack without any notch at (a)** *t* **= 0 ns, (b)** *t* **= 10 ns, and at the end of the racetrack with a notch at (c)** *t* **= 0 ns, (d)** *t* **= 0.95 ns, (e)** *t* **= 0.975 ns, (f)** *t* **= 1 ns (g)** *t* **= 1.025 ns (h)** *t* **= 1.05 ns (i)** *t* **= 1.075 ns (j)** *t* **= 1.1 ns (k)** *t* **= 1.125 ns and (l)** *t* **= 1.15 ns.** The simulation parameters are the same as in Fig. 5.

In domain wall-based RM, domain-wall bits are annihilated simply by pushing them with driving currents, beyond the reading element, and eventually they will be pushed out of the racetrack by these currents [27]. In the same way, once the skyrmionic bits in skyrmion-based RM are pushed beyond the reading element they can no longer be accessed and so should be eliminated at the end of the racetrack. To prove the feasibility, we performed simulations of skyrmionic bit chain moving toward the end of the racetrack.

Figure 6 shows the vertical-current-driven motion of a skyrmionic bit chain with $d_i = 60$ nm at the end of 40-nm-wide racetracks, with the same driving current as in Fig. 5. As shown in Fig. 6a and 6b, because of the skyrmion-skyrmion and skyrmion-edge repulsions, the skyrmionic bits experience congestion at the end of the racetrack: the skyrmionic bits driven by spin currents are moving steadily toward the end of the racetrack, but when the first skyrmionic bit approaches the edge, it is slowed down due to the repulsions. The spacings between the skyrmionic bits then reduce significantly, and at the same time, the skyrmion size $r_s$ decreases (see Supplementary Movie 3). For the head skyrmionic bit of the chain as shown in Fig. 6b, $r_s$ can thus be compressed from 10 nm to 7 nm due to the repulsive forces from edges and neighbor skyrmionic bits. At this current density (~ $10^{10}$ A/m²), the skyrmions cannot easily exit from the track. If the current density is increased to be larger than a critical threshold (~ $10^{11}$ A/m² in our case), the skyrmionic bits can exit the racetrack because the driving force from currents can overcome the repulsive force from the edge at the end of the racetrack, as shown in Ref. 19.

However, because one important design concept of skyrmion-based RM is the extremely lower power-consumption, we would like a method to avoid the observed clogging of skyrmions on racetracks that does not require increasing the current density. As shown in Fig. 6c-l, we propose a triangular notch with rounded tip at the end of the racetrack with depth and width respectively equal to 34 nm and 40 nm, which can be realized by nanolithography [36, 37] and provides a design to resolve the issue. When moving skyrmionic bits touch the notch-tip, they will soon be eliminated (see Supplementary Movie 4). Figure 6d-l show the process of the annihilation of a skyrmion within 0.2 ns, where the skyrmionic bit is firstly attracted by the tilts of magnetization around the notch-tip, then evolves to a domain wall pair [22] when touching the notch edge and is eventually cleared away by the currents. In this process, the skyrmionic bit chain is not compressed at the end of the racetrack, in which all skyrmions move together coherently.

## Discussion

In order to realize reliable and controlled motion of bits on racetrack, the domain wall-based RM uses a series of "pinning sites" along the racetrack, *i.e.*, preferred positions for the domain walls along the track where the energy of the domain wall is decreased. These "pinning sites" can play an important role in defining energetically stable positions for the domain walls and thereby the spacing between consecutive bits [27, 29]. The skyrmions in skyrmion-based RM should also keep on an appropriate spacing to avoid the bit interference and write/read errors.

In Ref. 20, Sampaio *et al.* report that the skyrmionic bit chain on the racetrack moves as single isolated skyrmionic bit where the initial spacing between consecutive skyrmionic bits is ~ 60 nm. However, skyrmions repel each other due to the effective skyrmion-skyrmion interaction, which according to the model in Ref. 21 decays exponentially with the skyrmion separation distance. Hence a reliable and practicable initial spacing is required for the consecutive skyrmionic bits in skyrmion-based RM, which ensures that the skyrmionic bits are not significantly influenced by their mutual repulsion, as shown in Fig. 2. So that the storage density, processing speed and data robustness will not be affected.

Obviously, the storage density of skyrmion-based RM and the bit identification by reading element will depend on the value of equilibrium distance $d_e$ between consecutive skyrmionic bits. In order to realize an ultra-high density, the equilibrium distance $d_e$ shall be as small as possible. It can be seen from Fig. 3, both equilibrium distance $d_e$ and relaxed skyrmion size $r_s$ decrease approximately linearly with increasing PMA and increasing applied field. $d_e$ and $r_s$ also change with the width of racetrack.

The applied field perpendicular to the racetrack plane and the PMA have the same physical effect: considering the fabricating cost and feasibility, a nanotrack with appropriately high PMA would be a better choice for building skyrmion-based RM than a static applied field.

Figure 4 shows increasing the track width generally results in an increased equilibrium distance $d_e$ as well as an increase in the relaxed skyrmion size $r_s$. From the point of view of storage density, within the range of track width with which the skyrmion is stable, the narrow width of the track offers possibility of fabricating high-density racetrack memory device.

As show in Fig. 5, only when the spacing is larger than a certain threshold (~ 60 nm in our case), the skyrmionic bit chain is not influenced by repulsions. Hence it can be seen that a skyrmionic bit chain can move as a static coherent unit when the skyrmion-skyrmion repulsion is small enough ($F_{ss}$ $\approx 10^{-12}$ N/m), and correspondingly the skyrmion spacing is sufficiently large. Interestingly, it should be mentioned that the repulsion between domain walls can enable denser domain wall-based RM, as shown in Ref. 38.

On the other hand, as shown in Fig. 6, we found that the skyrmionic bit chain will get clogged at the end of normal racetrack and the size of skyrmions will reduce, which is caused by the repulsions between skyrmions and edges as well as the repulsions between skyrmions and skyrmions. We therefore propose a technological solution to address this problem and demonstrate that the end of the racetrack can be designed with notch-tip to ensure that the skyrmionic bits moving beyond the reading element can easily exit from the racetrack at low current density ($\sim 10^{10}$ A/m$^2$). Due to the Dzyaloshinskii-Moriya interaction, the magnetization around the notch-tip is more tilted than that around the flat edge, *i.e.*, the magnetization around the notch-tip has a more dominant in-plane component, which facilitates the easier reversal of the skyrmion boundary as well as the core of skyrmion while it approaches the notch-tip.

To summarize, our data have clearly demonstrated that skyrmion-skyrmion and skyrmion-edge repulsions exist and play an important role in skyrmion-based magnetic racetrack memory. The reliable and practicable spacing between consecutive skyrmionic bits on skyrmion-based racetrack memory has been studied, which is important for the writing process, reading process and storage density. Besides, we have tested that the reliable and practicable spacing can be adjusted in a limited range by tuning the perpendicular anisotropy and/or applying an external magnetic field perpendicular to the racetrack plane. Finally, we conclude that the notch-tip is a practical way to assist skyrmionic bits moving beyond the reading element to exit from the racetrack easily. This can effectively avoid clogging, compression and congestion of skyrmionic bits on the racetrack. Our results are relevant to the fundamental understanding of magnetic skyrmions, and more practically, to the application of magnetic skyrmions in memory technology.

## Methods

**Micromagnetic simulations.** The micromagnetic simulations are performed using the software package OOMMF [39] including the extension module of the Dzyaloshinskii-Moriya interaction (DMI) [20, 40, 41]. The motion of skyrmionic bits is simulated in 0.4-nm-thick cobalt racetracks with width of

30 ~ 100 nm and length of 200 ~ 800 nm.

The simulation involving magnetization dynamics without a spin transfer torque relies on the Landau-Lifshitz-Gilbert (LLG) ordinary differential equation [42-45],

$$\frac{d\mathbf{M}}{dt} = -|\gamma|\mathbf{M} \times \mathbf{H}_{\text{eff}} + \frac{\alpha}{M_S}\left(\mathbf{M} \times \frac{d\mathbf{M}}{dt}\right), \qquad (1)$$

where $\mathbf{M}$ is the magnetization, $\mathbf{H}_{\text{eff}}$ is the effective field, $\gamma$ is the Gilbert gyromagnetic ration, and $\alpha$ is the damping coefficient. The effective field is defined as follows:

$$\mathbf{H}_{\text{eff}} = -\mu_0^{-1}\frac{\partial E}{\partial \mathbf{M}}. \qquad (2)$$

The average energy density $E$ is a function of $\mathbf{M}$ specified by,

$$E = A\left[\nabla\left(\frac{\mathbf{M}}{M_S}\right)\right]^2 - K\frac{(\mathbf{n}\cdot\mathbf{M})^2}{M_S^2} - \mu_0\mathbf{M}\cdot\mathbf{H} - \frac{\mu_0}{2}\mathbf{M}\cdot\mathbf{H}_{\text{d}}(\mathbf{M}) + \omega_{\text{DM}}, \qquad (3)$$

where $A$ and $K$ are the exchange and anisotropy energy constants, respectively. $\mathbf{H}$ and $\mathbf{H}_{\text{d}}(\mathbf{M})$ are the applied and magnetostatic self-interaction fields while $M_S = |\mathbf{M}(\mathbf{r})|$ is the spontaneous magnetization. The $\omega_{\text{DM}}$ is the energy density of the DMI, which has the form [20, 45],

$$\omega_{\text{DM}} = \frac{D}{M_S^2}\left(M_z\frac{\partial M_x}{\partial x} - M_x\frac{\partial M_z}{\partial x} + M_z\frac{\partial M_y}{\partial y} - M_y\frac{\partial M_z}{\partial y}\right), \qquad (4)$$

where the $M_x$, $M_y$, $M_z$ are the components of the magnetization $\mathbf{M}$ and $D$ is the DMI constant. The five terms at the right side of Equation (3) correspond to the exchange energy, the anisotropy energy, the applied field (Zeeman) energy, the magnetostatic (demagnetization) energy and the DMI energy, respectively.

For simulation with spin-polarized currents perpendicular to the racetrack plane (CPP), the in-plane and out-of-plane torques were added to the LLG equation, written as [20, 46]

$$\tau_{\text{IP}} = \frac{u}{t}\mathbf{m} \times \left(\mathbf{m}_p \times \mathbf{m}\right) \qquad (5)$$

$$\tau_{\text{OOP}} = -\xi\frac{u}{t}\left(\mathbf{m} \times \mathbf{m}_p\right) \qquad (6)$$

where $u = \gamma(\hbar jP/2\mu_0 eM_S)$, $\mathbf{m}$ is the reduced magnetization $\mathbf{M}/M_S$, $j$ is the current density, $P$ is the spin polarization, $\mathbf{m}_p$ is the unit electron polarization direction, $t$ is the film thickness and $\xi$ is the amplitude of the out-of-plane torque relative to the in-plane one. Thus, the LLG equation of magnetization dynamics becomes [39, 47],

$$\frac{d\mathbf{m}}{dt} = -|\gamma|\mathbf{m} \times \mathbf{H}_{\text{eff}} + \alpha\left(\mathbf{m} \times \frac{d\mathbf{m}}{dt}\right) + |\gamma|\beta\varepsilon\left(\mathbf{m} \times \mathbf{m}_p \times \mathbf{m}\right) - |\gamma|\beta\varepsilon'\mathbf{m} \times \mathbf{m}_p, \qquad (7)$$

where $\varepsilon = \frac{P\Lambda^2}{(\Lambda^2+1)+(\Lambda^2-1)(\mathbf{m}\cdot\mathbf{m}_p)}$, $\varepsilon' = P\xi/2$ and $\beta = \left|\frac{\hbar}{\mu_0 e}\right|\frac{j}{tM_S}$. Note that we have set $\Lambda = 1$ and $\xi =$ 0 to remove the dependence of $\varepsilon$ on $\mathbf{m} \cdot \mathbf{m}_p$ and the field-like out-of-plane torque respectively in all simulations.

The magnetic material parameters are adopted from Ref. 20: saturation magnetization $M_S$ = 580 kA/m, exchange stiffness $A$ = 15 pJ/m, DMI constant $D$ was fixed at 3 mJ/m$^2$, perpendicular magnetic anisotropy (PMA) $K$ = 0.8 MJ/m$^3$. Gilbert damping coefficient $\alpha$ = 0.3 and the value for $\gamma$ is -2.211×10$^5$. In all simulations with CPP, the electron polarization $P$ is fixed at 0.4 and the Oersted fields are included. A discretization of $1 \times 1 \times 0.4$ nm$^3$ is used in simulations of normal racetracks, while $0.5 \times 0.5 \times 0.4$ nm$^3$ is used in simulations of notched racetracks in order to ensure the numerical accuracy for considering the notch.

## Acknowledgments

The authors thank J. Sampaio & S. Rohart from Unité Mixte de Physique CNRS/Thales and Université Paris Sud for their useful suggestions in CPP & DMI. We also thank X.X. Liu from Shinshu University and J.J. Ding from CAS for helpful discussions. This work is supported by NSFC (Grant No. 11074179 and No. 10747007), the Construction Plan for Scientific Research Innovation Teams of Universities in Sichuan (No. 12TD008), and SSRIP of SICNU.

## Author contributions

G.P.Z., H.F. and X.C.Z. conceived and coordinated the project. X.C.Z. performed the micromagnetic simulations and analyzed the simulation data. X.C.Z., G.P.Z., H.F., J.P.L. and W.X.X. interpreted the results. X.C.Z., G.P.Z., H.F., J.X. and F.J.M. prepared the manuscript. All authors commented on the manuscript.

## Competing financial interests

The authors declare no competing financial interests.

## Supplementary information

Supplementary movies accompany this paper at http://www.nature.com/scientificreports

## How to cite this article

Zhang, X. C. *et al*. Skyrmion-skyrmion and skyrmion-edge repulsions in skyrmion-based racetrack memory. *Sci. Rep.* **5**, 7643; DOI: 10.1038/srep07643 (2015).